\documentclass[10pt,letterpaper]{article}
\usepackage[top=0.85in,left=2.75in,footskip=0.75in]{geometry}

\usepackage{amsmath,amssymb}

\usepackage{changepage}

\usepackage{textcomp,marvosym}

\usepackage{cite}

\usepackage{nameref,hyperref}

\usepackage[right]{lineno}

\usepackage[nopatch=eqnum]{microtype}
\DisableLigatures[f]{encoding = *, family = * }

\usepackage[table]{xcolor}

\usepackage{array}

\newcolumntype{+}{!{\vrule width 2pt}}

\newlength\savedwidth

\def\CPP{{C\nolinebreak[4]\hspace{-.05em}\raisebox{.4ex}{\tiny\bf ++}}}

\raggedright
\usepackage[aboveskip=1pt,labelfont=bf,labelsep=period,justification=raggedright,singlelinecheck=off]{caption}

\makeatletter
\renewcommand{\@biblabel}[1]{\quad#1.}
\makeatother

\usepackage{lastpage,fancyhdr,graphicx}
\usepackage{epstopdf}
\usepackage{siunitx,booktabs}
\fancyheadoffset[L]{2.25in}
\fancyfootoffset[L]{2.25in}
\begin{document}

\begin{flushleft}
{\Large
\textbf\newline{Biomedical open source software: Crucial packages and hidden heroes} 
}
\newline
\\
Eva Maxfield Brown\textsuperscript{1\Yinyang},
Stephan Druskat\textsuperscript{2\Yinyang},
Laurent H\'ebert-Dufresne\textsuperscript{3\Yinyang},
James Howison\textsuperscript{4\Yinyang*},
Daniel Mietchen\textsuperscript{5,6\Yinyang},
Andrew Nesbitt\textsuperscript{7\Yinyang},
João Felipe Pimentel\textsuperscript{8\Yinyang},
Boris Veytsman\textsuperscript{9,10\Yinyang}
\\
\bigskip
\textbf{1} Information School, University of Washington, Seattle, Washington, USA\\
\textbf{2} Institute of Software Technology, German Aerospace Center (DLR), Berlin, Germany\\
\textbf{3} Department of Computer Science, University of Vermont, Burlington, Virginia, USA\\
\textbf{4} School of Information, The University of Texas at Austin, Austin, Texas, USA\\
\textbf{5} Leibniz Institute for Information Infrastructure, FIZ~Karlsruhe, Berlin, Germany\\
\textbf{6} Institute for Globally Distributed Open Research and Education (IGDORE), Jena, Germany\\
\textbf{7} Ecosyste.ms, London, UK\\
\textbf{8} Instituto de Computação, Universidade Federal Fluminense, Niterói, Rio de Janeiro, Brazil\\
\textbf{9} Chan Zuckerberg Initiative, Redwood City, California, USA\\
\textbf{10} School of Systems Biology, George Mason University,
   Fairfax, Virginia, USA\\
\bigskip

%
%
\Yinyang These authors contributed equally to this work.

* jhowison@ischool.utexas.edu

\end{flushleft}

\begin{abstract}
  Despite the importance of scientific software for research, it is
  often not formally recognized and rewarded.  This is especially
  true for foundational libraries, which are hidden below packages visible to the users (and thus doubly hidden, since even the packages directly used in research are frequently not visible in the paper).  Research stakeholders like funders, infrastructure providers, and other organizations need to understand the complex
  network of computer programs that contemporary research relies upon.

  In this work, we use the CZ Software Mentions Dataset to map the
  upstream dependencies of software used in biomedical papers and find the
  packages critical to scientific software ecosystems.  We propose centrality
  metrics for the network of software dependencies, analyze three
  ecosystems (PyPi, CRAN, Bioconductor), and determine the packages
  with the highest centrality.
\end{abstract}

\section*{Author summary}

Scientists today use software as an important tool for their research.
The progress of science depends on the people who write software being
properly recognized, rewarded, and incentivised.  This means that those
who fund research or promote scientists need to know the impact of
software and identify the most important programs.

The universe of scientific software is complex: some people write
programs used by researchers, while some people write programs used by
other programs.  The latter are often even less visible than the former,
but they are important for the progress of science.

In this work, we establish metrics to measure the impact of both types
of software.  We introduce the network of dependencies (program~A used
by researchers depends on programs~B, C, and~D) and use it to
calculate the impact of a large number of packages used in the biomedical
literature.  In doing so, we identify important software packages, regardless of their direct visibility to end users.

We conclude with a discussion of the limitations of our methodology and approaches to improve network validity.

\section{Introduction}

Since the second half of the last century, a computer has become as
ubiquitous a tool of a scientific lab as balance scales and a Bunsen
burner were in previous ages.  As a consequence, computer software is
now crucial to research, bringing new methods and new scale, while
offering new potential for reproducibility and extension.  This is
true not only for the natural sciences and mathematics but for
scholarship more broadly (e.g., sociology and the humanities), making
the software revolution both wide and deep. Yet we have very limited
insight into the software actually used in research. This lack of
infrastructural understanding means we are limited in our ability to
reward developers and maintainers, encourage collaboration and
coordination, and direct science funding in a well-informed manner.

Scientific software is often invisible in publications, because
citation practices in science have not changed at the same pace as software has become crucial. For example, software is infrequently and
inconsistently formally cited~\cite{Howison2011, Howison2015,
  SinghChawla2016, Howison2016, Knowles2021, DruskatEtAl2024}. There
have been recent efforts to extract informal citations from the
full text of articles~\cite{Schindler2021, Du2021,
  SoftwareMentionsDataset2022, SoftwareMentionsArXiv2022} and to
evaluate the ``importance'' of software packages by looking at papers
that cite them~\cite{Bogart2020}.  Unfortunately, publications
sometimes do not mention all the software used in the course of research.

Besides this, there is another kind of invisibility of scientific
software.  The programs visible to the end user may rely on many other
software packages (known in the software world as dependencies). While
the end users may mention a package at the top of the dependency
stack, they are likely not even aware of the full set of packages that are further
below. These may be packages that the user-facing program depends on directly
(\emph{direct dependencies}), or indirectly, where the direct
dependencies in turn may depend on other
packages. These latter packages thus become indirect, \emph{transitive
  dependencies} of the user-facing program.  This complex nature of
the network of dependencies has a number of implications, including those
for security~\cite{Hatta2022Nebraska,Goodin2024} and computational
reproducibility~\cite{Samuel2024Computational}.  In particular, much of
the work undertaken to develop, maintain, test, and distribute the
underlying software is not directly visible in the publication record
itself, and is not included in derivatives such as citation networks and
knowledge graphs.

The situation resembles the famous XKCD
cartoon where ``all modern infrastructure''
critically depends on ``a project some random person in Nebraska has
been thanklessly maintaining since 2003''~\cite{Munroe2020}.  The word ``thanklessly''
is important in this context: being unknown, these critical pieces of
software get much less recognition and credit than they deserve---and
than the science needs.  The absence of recognition may lead to dire
consequences, for example, if, due to the lack of funded maintenance, the
underlying libraries become vectors for malware attacks.  

There have been proposals to assign credit to these packages using the dependency structure to calculate what has been termed \emph{transitive credit}~\cite{Katz2014,KatzSmith2014} but so far, the problem has not yet been solved. To implement the proposed  measure, software projects must start to publicize the packages they rely on in a way
that enables recognition and citation, beyond the technical dependency
already recorded in manifest files such as \texttt{pyproject.toml},
\texttt{DESCRIPTION}, \texttt{cargo.toml}, \texttt{pom.xml}, etc. One
way for projects to do this is the inclusion of citation information
for their own software outputs as well as for their direct
dependencies, e.g., a citation file in the Citation File Format
(CFF)~\cite{DruskatCitationFileFormat2021}.  Were such citation
information available for the complete dependency stack of a program,
transitive credit could be implemented by building weighted software
citation networks~\cite{DruskatSoftwareDependenciesCitationGraphs2020}.

At present, though, the situation is quite different: sometimes even the maintainers of the lower-stack software are not aware of the upper-stack programs that depend on their work.  They need this knowledge when making breaking changes to their packages, which might negatively influence the software that depends on them~\cite{Bogart2021}.

By classifying software packages as visible to the end users or
primarily important for other packages, we follow the ideas described
by Donald E.~Stokes in his famous book, Pasteur’s
Quadrant~\cite{Stokes1997:PasteursQuadrant}.  There, Stokes
distinguishes between applied research with the results visible to the
general public, and pure research, which is less visible, but
important for the applied research.  He discussed a two-dimensional
diagram of scientific works, with the importance of the works for the
applications on the $x$~axis and the quest for fundamental
understanding on the~$y$ axis.  Thus, the upper left quadrant is
occupied by the purely fundamental research personified by Niels Bohr.
The lower right quadrant is occupied by purely applied research
personified by Thomas Edison.  The upper right quadrant is occupied by
the works that have both fundamental and practical value.  Stokes
chose Louis Pasteur as an example of such works.  While this picture
is rather simplified (Bohr's works have significant practical value,
and Edison's experiments stimulated fundamental research), it gives
important insight into scientific endeavor.

\begin{figure}
  \centering
  \includegraphics{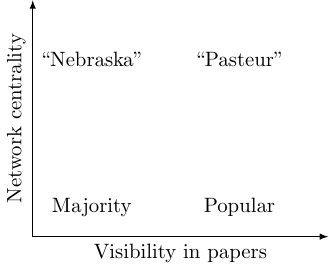}
  \caption{Classification of software packages inspired by Stokes' classification system in ~\cite{Stokes1997:PasteursQuadrant}. “Nebraska” packages are software projects which have few mentions in research articles, but are highly central in a dependency network. “Pasteur” packages are both highly visible with lots of mentions and are highly central in a dependency network.}
  \label{fig:pasteur}
\end{figure}

Following these ideas, we can put the software packages into a two-dimensional plane with the axes corresponding to the frequency of
software mentions and a measure of network centrality
(\autoref{fig:pasteur}).  The majority of the packages will likely occupy the
lower left corner of the plot, having a small number of mentions from the authors of scientific papers and limited centrality from usage in other software packages. Popular packages are used by many authors.  We are interested in the ``Nebraska'' packages, not very visible, but of critical importance. We use the term ``Nebraska'' packages throughout this article in reference to the original XKCD cartoon which noted that much of highly critical software is often less visible and less well known~\cite{Munroe2020}.  Of course, one can think about highly visible and critical ``Pasteur'' packages, which are used both directly and as a foundation for other libraries.

We had a fortunate opportunity to explore these ideas due to the
generosity of the Chan Zuckerberg Initiative (CZI).  On 24--27 October
2023, CZI hosted a hackathon on \textit{Mapping the Impact of Research
  Software in Science} (see
\url{https://github.com/chanzuckerberg/software-impact-hackathon-2023}).
One of the projects at this hackathon was \emph{Tracing the
  dependencies of open source software mentioned in the biomedical
  literature}~\cite{Brown_Exploring_the_dependencies_2023}.  In this
project, we explored the dependencies of the open source software
packages mentioned in the CZ Software Mentions
dataset~\cite{SoftwareMentionsDataset2022}.  The dataset was created using a machine learning system trained on the SoftCite gold standard dataset of manually annotated software mentions~\cite{Du2021} and consists of extractions from 2.4 million biomedical papers. The methodology and its
limitations are discussed in~\cite{SoftwareMentionsArXiv2022}.  In
particular, the work on the disambiguation of packages can be improved.
Later, we discuss how these limitations influence our results.  

We decided to limit our
study to open-source packages: first, because the dependencies of
closed-source packages are not public, and second, because we believe
in the importance of open source for open and reproducible science. Here, we explore approaches
to making the software package infrastructure underlying science more
visible. We examine our findings to develop questions to better
understand the idea of criticality and opportunities for improving
science through adjustments to the research software ecosystems.

\section{Materials and Methods}

\subsection{Network construction}
\label{sec:construction}

We combined three datasets: The CZ Software Mentions
dataset~\cite{SoftwareMentionsDataset2022}, which was constructed using a
machine learning model trained on the SoftCite Gold Standard
annotations~\cite{Du2021}, the Ecosyste.ms software dependency
dataset~\cite{Nesbitt2023Data}, which is built by gathering and
normalizing the dependency information from multiple software ecosystems,
including one for the Python programming language (PyPI) and two for the R programming language (CRAN and Bioconductor), and finally we used 
  OpenAlex~\cite{OpenAlex} for information on how frequently papers had been cited. The software authorship data were scraped from GitHub metadata.

The CZ Software Mentions dataset~\cite{SoftwareMentionsDataset2022,
  SoftwareMentionsArXiv2022} includes mentions from 2.4 million biomedical papers and identifies which mentions were traced to
which ecosystem. In this way, package names from CRAN, Bioconductor, and PyPI are parsed by the software mentions extraction model, collected, and finally disambiguated. We collected information for each package, including its dependencies, using the latest release of the package at the time of data
collection (October 2023). Dependencies were then recursively
retrieved using the most recent release and following dependencies until the full
list of transitive dependencies was obtained. Limitations of our
dependency resolution approach are discussed later.

After this data processing, we produced a two-mode network, with nodes
for papers and nodes for software packages. Edges from papers to software packages were
added when a package was mentioned in the full text of the
article. Edges between packages were added when metadata
descriptions indicated a required dependency.  The full code for the processing is available at
\cite{Brown_Exploring_the_dependencies_2023}.  The network is
available at~\cite{Brown_dataset} in GEXF format~\cite{GEXF} and 
has four classes of nodes:
\begin{description}
\item[paper:] papers from the CZ Software Mentions dataset.  They are
  identified by their Digital Object Identifier (DOI). To estimate the impact of the papers, we
  separately downloaded their citation numbers from
  OpenAlex~\cite{OpenAlex} as of November 2023 (a copy is available in
  the \path{data} subdirectory of~\cite{Brown_Exploring_the_dependencies_2023}). 
\item[pypi, cran, bioconductor:] Software from the corresponding
  ecosystems.  We used CZI ID~\cite{SoftwareMentionsDataset2022,
    SoftwareMentionsArXiv2022} as the identifier.  We did not attempt
  to identify the same software across the ecosystems (see the
  discussion below). 
\end{description}
The edges are directed and weighted.  An edge from one software node~$A$
to another software node~$B$ means that software~$A$ depends on software~$B$
as determined by the corresponding metadata.  An edge from a
paper node to a software node means that the given software is
mentioned in the paper, and the weight corresponds to the number of
citations the paper received.  In what follows, we will consider both
the weighted and unweighted versions of this network, as utilization of the weights has both positive and negative aspects to consider during analysis and interpretation.

\subsection{Network analysis}

To analyze the network, we relied on directed centrality
analysis~\cite{Borgatti2021}. There are several possible options for
centrality measures. Our choice was determined by the following
considerations. First, we wanted a centrality metric that can account
for papers even if they have no incoming edges in our network (in
other words, papers do not receive centrality but should contribute to
the centrality of software packages). This criterion excludes
eigenvector centrality~\cite{Bonacich2007Some}. Second, we wanted a
centrality metric that can account for the weight of edges from papers
to software packages (i.e., some papers contribute more centrality
because they are more highly cited.). This criterion excludes the PageRank
algorithm~\cite{Brin1998anatomy}, which normalizes weights of
out-degree. We found that Katz centrality~\cite{Katz1953new} satisfies
all our criteria. Katz
centrality can be interpreted as the importance of the node for
the diffusion processes on the network.  This corresponds to the
intuition that innovation is a diffusion-like
process~\cite{Rogers2003diffusion}. Katz centrality gives us the
attenuation factor $\beta$ as a free parameter to control the
importance of papers (which we set equal to 1), such that paper nodes
contribute a factor proportional to $\beta$ times their citation count
to the software packages they mention (we follow the formulation of
Katz centrality in~\cite{Sun2011}). In turn, software packages
contribute a factor proportional of their own centrality to their
dependencies. A package can therefore be central by receiving mentions
from well-cited papers, by having central dependents, or by a
combination of both.

We additionally provide analysis for three variations of our network: unweighted, weighted, and the largest connected component from the weighted graph. The unweighted form of our network ignores the weights added to the edges between the seeding articles and their software mentions. The weighted network includes the article citation count weights. The difference between the unweighted and weighted versions of our analysis, allow us to parse different subtypes of the results. Specifically, comparing the weighted and unweighted networks allows to identify software which is only important due to its utilization and reference in high citation count articles.

\begin{figure}
    \centering

    \includegraphics[width=0.9\linewidth]{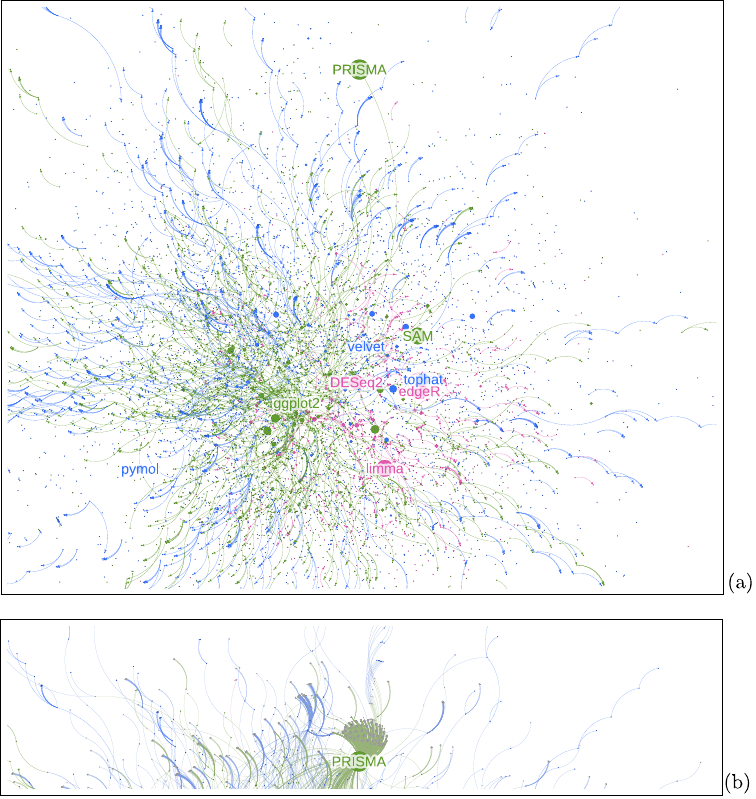}
     
    \caption{(a) Network visualization of software packages from three
      ecosystems (from CRAN in green, PyPI in blue, and Bioconductor
      in pink) connected through their dependencies within their
      ecosystem and interconnected through papers that mention
      them. We label the top 3 most central packages in each
      ecosystem: \texttt{ggplot2}~\cite{wickham2011ggplot2},
      \texttt{SAM}~\cite{ravikumar2009SAM}, 
      and \texttt{PRISMA}~\cite{krueger2012prisma} 
      for CRAN, \texttt{velvet}~\cite{velvet}, 
      \texttt{tophat} 
      and \texttt{pymol}~\cite{delano2002pymol} 
      for PyPI and \texttt{DeSeq2}~\cite{love2014deseq2},
      \texttt{edgeR}~\cite{robinson2010edger} 
      and \texttt{limma}~\cite{smyth2005limma,ritchie2015limma} 
      for Bioconductor. The core of the network is dominated by CRAN and PyPI
      dependencies, despite the fact that three of the five most central
      packages come from Bioconductor. (b) The top part of the above network,
      with papers added (in grey) to illustrate how  \texttt{PRISMA}~\cite{krueger2012prisma} can be
      central due to many mentions in  papers. }
    \label{fig:network-viz}
\end{figure}

\section{Results}
\label{sec:results}

In~\autoref{fig:network-viz}, we show the overall network of software
packages connected by their dependencies within each ecosystem
and interconnected through the papers that mention them. Edges from
papers to software packages are directed and weighted by the number of
citations the paper received. Edges between software dependencies are
directed from a dependent to a dependency.

We found a dense core of popular packages that receive many mentions
(e.g. \texttt{ggplot2} in CRAN~\cite{wickham2011ggplot2}, \texttt{tophat} in PyPI and
\texttt{limma} in Bioconductor~\cite{smyth2005limma,ritchie2015limma}), some of which have many dependencies
themselves (e.g. \texttt{ggplot2}~\cite{wickham2011ggplot2}). We also found packages specific to
certain communities (e.g. \texttt{PRISMA} in CRAN ~\cite{krueger2012prisma} or \texttt{pymol} in
PyPI~\cite{delano2002pymol}).

Analyzing the entire ecosystem of software dependencies, we found that
roughly 10\% of software packages are part of dependency loops (i.e.,
cycles in the dependency networks).  Interestingly, we found no cycles
in the connected components of networks that have received software
mentions, and their dependencies.  We discuss this finding below. Note
that contemporary package managers usually resolve the dependency loops by
installing all packages in the loop simultaneously, so the loops,
while indicating rather sloppy software practices, are not fatal.

\begin{figure}
  \centering
  \includegraphics[width=\linewidth]{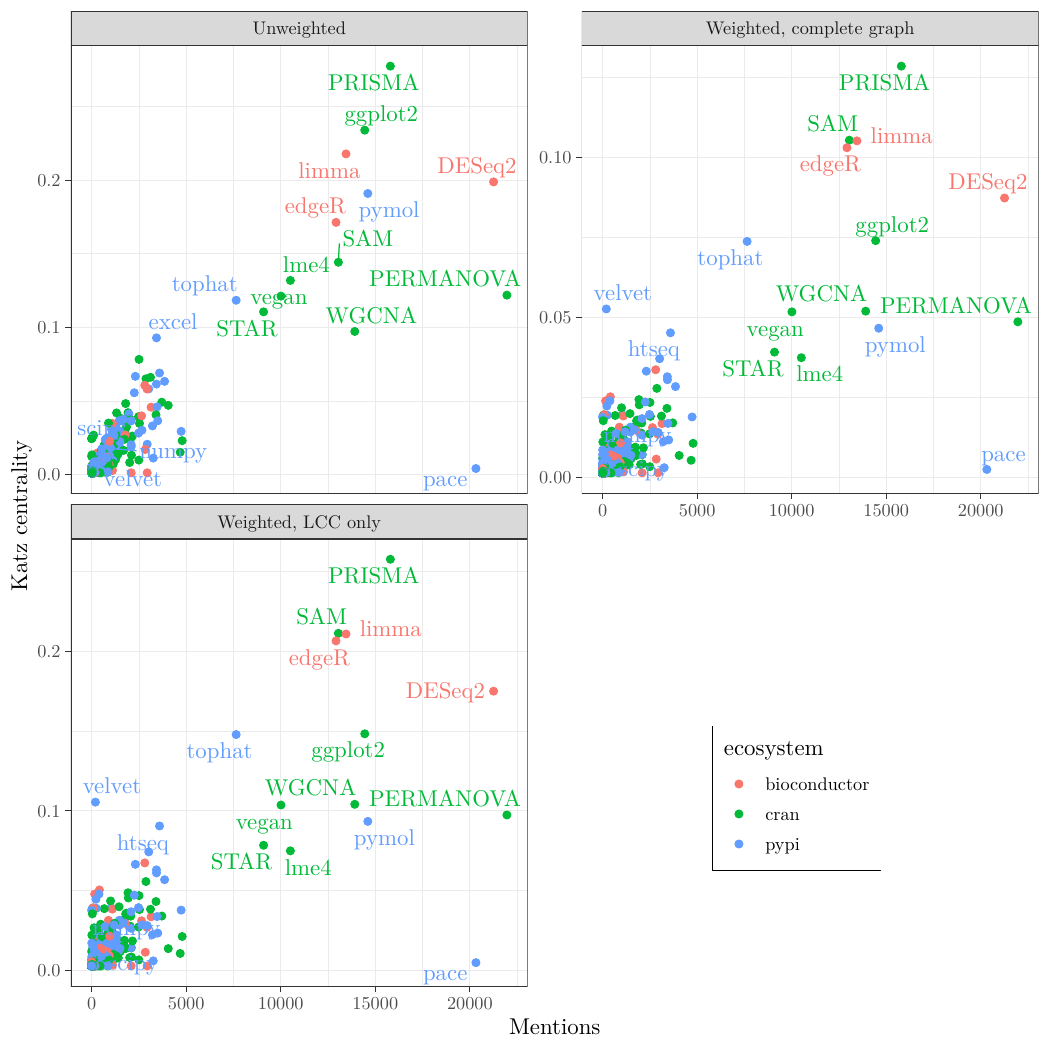}
  \caption{Distribution of packages by Katz centrality and counts of
    their mentions in papers.  Katz centrality is calculated for an
    unweighted graph, for a weighted graph with all nodes, or just for
    the largest connected cluster (LCC) for each ecosystem.  In the
    calculations, we assumed $\beta=1$.}
  \label{fig:results}
\end{figure}

In~\autoref{fig:results}, we show the distribution of packages from
the three investigated ecosystems over Katz centrality and mention
counts.  We used three different ways to calculate centrality:
(a)~unweighted graph, (b)~weighted complete graph, and (c)~the largest
connected components of the weighted graph (calculated separately for
each ecosystem). It is remarkable that all three models give rather
similar results for many packages.  In the next section, we discuss
the insights from the cases when the different methods do not agree.
As expected, we found a dense cluster of packages with low Katz
centrality and low mention counts, i.e., the ``Majority'' quadrant
(see~\autoref{fig:pasteur}).

\begin{table}
  \centering
  \caption{Summary statistics for the number of mentions in
    papers. ``Number'' is the number of packages, ``Dependency only''
    is the fraction of packages not mentioned in any papers and pulled
    by the dependencies only, ``Median'' is the median number of
    mentions, ``IQR'' is the interquartile range for mentions count, ``Max'' is the maximal number of
    mentions, and
    ``Gini'' is the Gini coefficient for the mentions count.}
  \label{tab:summary_mentions}
  \begin{tabular}{lS[table-format=4]S[table-format=1.2]*{2}{S[table-format=2]}S[table-format=5]S[table-format=1.2]}
    \toprule
    Ecosystem &  {Number} & {Dependency only} &
                                                \multicolumn{4}{c}{Mentions
                                                count}\\
    \cmidrule{4-7}
    & & & {Median} & {IQR} & {Max} & {Gini} \\
    \midrule
bioconductor & 1018 & 0.0471512770137525 & 24 & 68 &
21257 & 0.843712143943309\\
cran & 3594 & 0.11324429604897 & 9 & 34 & 21960 &
0.889196914901722\\
pypi & 5596 & 0.14349535382416 & 4 & 17 & 20322 &
0.888156736619396\\   
    \bottomrule
  \end{tabular}
\end{table}

\begin{table}
  \centering
  \caption{Summary statistics for Katz centrality. ``Median'' is the
    median centrality, ``IQR'' is the interquartile range, ``Max'' is
    the maximal centrality, and ``Gini'' is the Gini coefficient.}
  \label{tab:summary_centrality}
  \begin{tabular}{l*{2}{S[table-format=1.4]}*{2}{S[table-format=1.2]}}
    \toprule
    Ecosystem & {Median} & {IQR} & {Max} & {Gini}\\
    \midrule
    \multicolumn{5}{c}{\itshape Unweighted graph}\\
bioconductor & 0.00126366728535956 & 0.00065127467783916
& 0.217934004136626 & 0.549779090129221\\
cran & 0.00116646210956267 & 0.00048602587898445 &
0.277617982075916 & 0.496401393691949\\
pypi & 0.00108092155486141 & 0.00019441035159378 &
0.191008170440888 & 0.351314534735671\\
    \multicolumn{5}{c}{\itshape Weighted, complete graph}\\
    bioconductor & 0.00143155318286753 &
0.000322302811767757 & 0.105238137865536 &
0.358203102844196\\
cran & 0.00138817278338669 & 0.000147428626544985 &
0.128618817548222 & 0.253046882585352\\
pypi & 0.00138275023345159 & 0.00010438408625076 &
0.07377243622961 & 0.180193304141016\\
    \multicolumn{5}{c}{\itshape Weighted, LCC only}\\
    bioconductor & 0.00288301570993583 & 0.00075920092370992
& 0.210742476047381 & 0.371663779294751\\
cran & 0.00279343047601127 & 0.000369917791154937 &
0.257562976941226 & 0.280390754327007\\
pypi & 0.00278528636383631 & 0.00031286964272138 &
0.147731480149716 & 0.220244615942456\\    
    \bottomrule
  \end{tabular}
\end{table}

Figure~\ref{fig:results} suggests large differences between the
packages in the majority quadrant and the packages in the rest of the
system.  This is further illustrated by
Tables~\ref{tab:summary_mentions} and~\ref{tab:summary_centrality}.
The sum of the median and half of the interquartile range is several
orders of magnitude smaller than the maximal value for both the number
of mentions and centrality.

A convenient measure of this inequality is the Gini
coefficient~\cite{Gini1909}.  Gini coefficients close to one
characterize a very unequal distribution, while Gini coefficients
close to zero correspond to a uniform distribution.  As seen from
Table~\ref{tab:summary_mentions}, the distribution of mention counts
is wildly unequal with high Gini coefficients.  The distribution of
centrality (Table~\ref{tab:summary_centrality}) is smoother,
especially for the weighted graph, but is still rather unequal.

Perhaps an even more robust measure of ``Nebraska'' properties of the
ecosystem is the number of packages that are not mentioned in any
paper, but are in the dependencies of the packages that are mentioned
(we are grateful to the anonymous reviewer of PLOS Computational Biology for
suggesting this metric).  It is shown in
Table~\ref{tab:summary_mentions}.

In Tables~\ref{tab:unweighted}, \ref{tab:weighted},
and~\ref{tab:weighted_LCC}, we show the packages in ``Pasteur'',
Popular, and ``Nebraska'' packages using the following criteria:
\begin{enumerate}
\item ``Pasteur'': the highest sum of mentions and centrality
  percentile.  To provide a ``sanity check'', we added the data for
  three important packages: \texttt{numpy}, \texttt{scipy}, and
  \texttt{napari}. 
\item Popular: the highest mentions percentile while centrality
  percentile is less than $0.5$.
\item ``Nebraska'': the highest centrality percentile while mentions
  percentile is less than $0.5$.
\end{enumerate}
These metrics calculated for all packages are available as CSV files in the
repository~\cite{Brown_Exploring_the_dependencies_2023}
(\url{https://github.com/borisveytsman/SoftwareImpactHackathon2023_Tracing_dependencies/tree/main/data}). 

\begin{table}
  \centering
\caption{Software quadrants, unweighted graph}                                       
  \label{tab:unweighted}
\begin{tabular}{llS[table-format=5]S[table-format=1.4]S[table-format=1.5]S[table-format=1.4]}
  \toprule
  Software & Ecosystem & \multicolumn{2}{c}{Mentions} & \multicolumn{2}{c}{Centrality}\\
  \cmidrule{3-6}
           & & {Count} & {Percentile} & {Value} & {Percentile}\\
  \midrule
\multicolumn{6}{c}{\bfseries``Pasteur''}\\
PRISMA & cran & 15797 & 0.9997 & 0.27762 & 1\\
DESeq2 & bioconductor & 21257 & 0.9999 & 0.19888 &
0.9997\\
ggplot2 & cran & 14441 & 0.9995 & 0.23407 & 0.9999\\
pymol & pypi & 14604 & 0.9996 & 0.19101 & 0.9996\\
PERMANOVA & cran & 21960 & 1 & 0.1219 & 0.9992\\
limma & bioconductor & 13451 & 0.9993 & 0.21793 &
0.9998\\
SAM & cran & 13048 & 0.9992 & 0.14425 & 0.9994\\
edgeR & bioconductor & 12923 & 0.9991 & 0.17137 &
0.9995\\
lme4 & cran & 10510 & 0.999 & 0.13191 & 0.9993\\
WGCNA & cran & 13915 & 0.9994 & 0.09721 & 0.9988\\
vegan & cran & 10012 & 0.9989 & 0.12121 & 0.9991\\
tophat & pypi & 7642 & 0.9987 & 0.1184 & 0.999\\
STAR & cran & 9091 & 0.9988 & 0.11052 & 0.9989\\
\multicolumn{6}{c}{\dotfill}\\
scipy & pypi & 1318 & 0.9921 & 0.02081 & 0.9916\\
numpy & pypi & 962 & 0.9884 & 0.0176 & 0.99\\
\multicolumn{6}{c}{\dotfill}\\
napari & pypi & 11 & 0.5873 & 0.00118 & 0.6104\\
\multicolumn{6}{c}{\bfseries Popular}\\
GSVA & bioconductor & 2937 & 0.9969 & 0.00108 & 0.361\\
MAST & bioconductor & 2091 & 0.995 & 0.00108 & 0.361\\
FSA & cran & 499 & 0.9782 & 0.00108 & 0.361\\
Boruta & cran & 441 & 0.9754 & 0.00108 & 0.361\\
MixSIAR & cran & 377 & 0.9714 & 0.00117 & 0.4989\\
DRIMSeq & bioconductor & 335 & 0.9678 & 0.00117 &
0.4989\\
netDx & bioconductor & 174 & 0.9419 & 0.00107 & 0\\
COCOA & bioconductor & 162 & 0.9378 & 0.00117 & 0.4989\\
emoji & pypi & 161 & 0.9374 & 0.00107 & 0\\
aviso & pypi & 160 & 0.937 & 0.00117 & 0.4989\\
SiZer & cran & 159 & 0.9366 & 0.00117 & 0.4989\\
searchlight & pypi & 137 & 0.9273 & 0.00107 & 0\\
\multicolumn{6}{c}{\bfseries``Nebraska''}\\
vctrs & cran & 2 & 0.2516 & 0.02438 & 0.9928\\
withr & cran & 2 & 0.2516 & 0.02438 & 0.9928\\
isoband & cran & 0 & 0 & 0.02438 & 0.9928\\
gss & cran & 7 & 0.4991 & 0.01212 & 0.9849\\
SuppDists & cran & 3 & 0.3287 & 0.00567 & 0.9658\\
lfda & cran & 1 & 0.1232 & 0.0049 & 0.9581\\
ggtext & cran & 7 & 0.4991 & 0.00482 & 0.957\\
texttable & pypi & 1 & 0.1232 & 0.00478 & 0.9569\\
affyio & bioconductor & 5 & 0.4326 & 0.00473 & 0.956\\
dnaio & pypi & 1 & 0.1232 & 0.00472 & 0.9558\\
requests & pypi & 7 & 0.4991 & 0.00419 & 0.9499\\
tifffile & pypi & 4 & 0.3863 & 0.00409 & 0.9488\\
  \bottomrule
\end{tabular}
  \end{table}

\begin{table}
  \centering
\caption{Software quadrants, weighted graph (full)}                                       
  \label{tab:weighted}
\begin{tabular}{llS[table-format=5]S[table-format=1.4]S[table-format=1.5]S[table-format=1.4]}
  \toprule
  Software & Ecosystem & \multicolumn{2}{c}{Mentions} & \multicolumn{2}{c}{Centrality}\\
  \cmidrule{3-6}
           & & {Count} & {Percentile} & {Value} & {Percentile}\\
  \midrule
\multicolumn{6}{c}{\bfseries``Pasteur''}\\
PRISMA & cran & 15797 & 0.9997 & 0.12862 & 1\\
DESeq2 & bioconductor & 21257 & 0.9999 & 0.08735 &
0.9996\\
SAM & cran & 13048 & 0.9992 & 0.10545 & 0.9999\\
limma & bioconductor & 13451 & 0.9993 & 0.10524 &
0.9998\\
ggplot2 & cran & 14441 & 0.9995 & 0.07401 & 0.9995\\
PERMANOVA & cran & 21960 & 1 & 0.04861 & 0.999\\
edgeR & bioconductor & 12923 & 0.9991 & 0.10309 &
0.9997\\
WGCNA & cran & 13915 & 0.9994 & 0.05195 & 0.9992\\
pymol & pypi & 14604 & 0.9996 & 0.04661 & 0.9989\\
tophat & pypi & 7642 & 0.9987 & 0.07377 & 0.9994\\
vegan & cran & 10012 & 0.9989 & 0.05172 & 0.9991\\
lme4 & cran & 10510 & 0.999 & 0.03738 & 0.9986\\
\multicolumn{6}{c}{\dotfill}\\
scipy & pypi & 1318 & 0.9921 & 0.00762 & 0.9881\\
numpy & pypi & 962 & 0.9884 & 0.0083 & 0.9891\\
\multicolumn{6}{c}{\dotfill}\\
napari & pypi & 11 & 0.5873 & 0.00136 & 0.2967\\
\multicolumn{6}{c}{\bfseries Popular}\\
GSVA & bioconductor & 2937 & 0.9969 & 0.00136 & 0.1214\\
MAST & bioconductor & 2091 & 0.995 & 0.00136 & 0.1214\\
GSA & cran & 844 & 0.9868 & 0.00136 & 0.2163\\
FSA & cran & 499 & 0.9782 & 0.00136 & 0.1508\\
Boruta & cran & 441 & 0.9754 & 0.00136 & 0.0968\\
MixSIAR & cran & 377 & 0.9714 & 0.00138 & 0.4265\\
AntWeb & cran & 282 & 0.9616 & 0.00138 & 0.4265\\
emoji & pypi & 161 & 0.9374 & 0.00138 & 0.4356\\
eureqa & pypi & 149 & 0.9331 & 0.00138 & 0.4265\\
searchlight & pypi & 137 & 0.9273 & 0.00136 & 0.259\\
pathfindR & cran & 133 & 0.9259 & 0.00137 & 0.4023\\
ChIPXpress & bioconductor & 125 & 0.9213 & 0.00138 &
                                                        0.4129\\
  pypet & pypi & 125 & 0.9213 & 0.00136 & 0\\
ADAPTS & cran & 125 & 0.9213 & 0.00137 & 0.3452\\
\multicolumn{6}{c}{\bfseries``Nebraska''}\\
calculate\_expression & pypi & 3 & 0.3287 & 0.01872 &
0.9956\\
zscore & pypi & 5 & 0.4326 & 0.00851 & 0.9893\\
webpages & pypi & 2 & 0.2516 & 0.00707 & 0.9872\\
setuptools & pypi & 6 & 0.4689 & 0.00572 & 0.9825\\
gpviz & pypi & 1 & 0.1232 & 0.00453 & 0.9759\\
glib & pypi & 3 & 0.3287 & 0.00397 & 0.9711\\
ego & pypi & 5 & 0.4326 & 0.00326 & 0.9607\\
xtick & pypi & 1 & 0.1232 & 0.00309 & 0.9575\\
pauvre & pypi & 5 & 0.4326 & 0.00292 & 0.9531\\
mappy & pypi & 4 & 0.3863 & 0.00292 & 0.9531\\
rle & cran & 2 & 0.2516 & 0.00287 & 0.9522\\
LineagePulse & bioconductor & 1 & 0.1232 & 0.0027 &
                                                       0.9474\\
  \bottomrule
\end{tabular}
  \end{table}

  \begin{table}
  \centering
\caption{Software quadrants, weighted graph (LCC component only)}                                       
  \label{tab:weighted_LCC}
\begin{tabular}{llS[table-format=5]S[table-format=1.4]S[table-format=1.4]S[table-format=1.4]}
  \toprule
  Software & Ecosystem & \multicolumn{2}{c}{Mentions} & \multicolumn{2}{c}{Centrality}\\
  \cmidrule{3-6}
           & & {Count} & {Percentile} & {Value} & {Percentile}\\
  \midrule
\multicolumn{6}{c}{\bfseries``Pasteur''}\\
PRISMA & cran & 15797 & 0.9996 & 0.25756 & 1\\
DESeq2 & bioconductor & 21257 & 0.9999 & 0.17492 &
0.9995\\
SAM & cran & 13048 & 0.999 & 0.21116 & 0.9999\\
limma & bioconductor & 13451 & 0.9991 & 0.21074 &
0.9998\\
ggplot2 & cran & 14441 & 0.9994 & 0.1482 & 0.9994\\
PERMANOVA & cran & 21960 & 1 & 0.09735 & 0.9988\\
edgeR & bioconductor & 12923 & 0.9989 & 0.20643 &
0.9996\\
  WGCNA & cran & 13915 & 0.9993 & 0.10402 & 0.999\\
  pymol & pypi & 14604 & 0.9995 & 0.09333 & 0.9986\\
tophat & pypi & 7642 & 0.9984 & 0.14773 & 0.9993\\
vegan & cran & 10012 & 0.9986 & 0.10357 & 0.9989\\
lme4 & cran & 10510 & 0.9988 & 0.07485 & 0.9983\\
\multicolumn{6}{c}{\dotfill}\\
scipy & pypi & 1318 & 0.99 & 0.01525 & 0.9851\\
numpy & pypi & 962 & 0.9854 & 0.01662 & 0.9863\\
\multicolumn{6}{c}{\dotfill}\\
napari & pypi & 11 & 0.5351 & 0.00273 & 0.2816\\
\multicolumn{6}{c}{\bfseries Popular}\\
GSVA & bioconductor & 2937 & 0.996 & 0.00272 & 0.1185\\
MAST & bioconductor & 2091 & 0.9937 & 0.00272 & 0.1185\\
GSA & cran & 844 & 0.9833 & 0.00272 & 0.2167\\
FSA & cran & 499 & 0.9724 & 0.00272 & 0.148\\
Boruta & cran & 441 & 0.969 & 0.00272 & 0.0923\\
MixSIAR & cran & 377 & 0.9639 & 0.00276 & 0.3878\\
AntWeb & cran & 282 & 0.9516 & 0.00276 & 0.3878\\
emoji & pypi & 161 & 0.9215 & 0.00276 & 0.397\\
searchlight & pypi & 137 & 0.9091 & 0.00273 & 0.254\\
pathfindR & cran & 133 & 0.9073 & 0.00275 & 0.3677\\
ChIPXpress & bioconductor & 125 & 0.9018 & 0.00276 &
0.3765\\
pypet & pypi & 125 & 0.9018 & 0.00272 & 0\\
\multicolumn{6}{c}{\bfseries``Nebraska''}\\
calculate\_expression & pypi & 3 & 0.2982 & 0.0375 &
0.9944\\
zscore & pypi & 5 & 0.3883 & 0.01705 & 0.9865\\
setuptools & pypi & 6 & 0.4215 & 0.01146 & 0.978\\
newick & pypi & 9 & 0.4957 & 0.00806 & 0.9649\\
xtick & pypi & 1 & 0.1305 & 0.00618 & 0.9469\\
nanolyse & pypi & 9 & 0.4957 & 0.00595 & 0.9443\\
splatter & pypi & 8 & 0.4742 & 0.00584 & 0.9422\\
pauvre & pypi & 5 & 0.3883 & 0.00584 & 0.9413\\
mappy & pypi & 4 & 0.3474 & 0.00584 & 0.9413\\
rle & cran & 2 & 0.232 & 0.00574 & 0.9402\\
scone & pypi & 8 & 0.4742 & 0.00574 & 0.9398\\
LineagePulse & bioconductor & 1 & 0.1305 & 0.0054 &
                                                       0.9344\\
  \bottomrule
\end{tabular}
  \end{table}

\section{Discussion}

Generally, the results verify our intuition that the majority of
software packages will have a low Katz centrality and relatively few
mentions.  The outliers are then the most interesting packages,
because they are central and often mentioned (``Pasteur''), or because
their centrality outweighs their mentions (``Nebraska''), or vice
versa (popular packages).

The absence of cycles in the network of open-source software packages
used in our sample of biomedical science is quite interesting.  It
suggests a more robust design structure in the universe of scientific
software than in the general software world. The lack of cycles also
improves our analysis with Katz centrality, as the absence of loops
excludes feedback mechanisms that can artificially inflate the
centrality of packages on a loop. Packages with low mentions and high
Katz centrality are therefore even more critical, as they represent
essential dependencies of packages that enable large volumes of
research without receiving direct attention.  These are the
``Nebraskan'' packages referenced by~\cite{Munroe2020} that we set out
to find.  \autoref{fig:results} and Tables~\ref{tab:unweighted},
\ref{tab:weighted}, and~\ref{tab:weighted_LCC} provide examples of
these packages.  As expected, most of these software packages are libraries used
by other packages: \texttt{vctrs}~\cite{vctrs} (a package for type
manipulation in R), \texttt{withr}~\cite{withr} (a set of tools for
safely running functions that change global variables),
\texttt{isoband}~\cite{isoband} (a package to generate isolines from
gridded data),
\texttt{pauvre}~\cite{pauvre} (a package for plotting Oxford Nanopore data), \texttt{newick}~\cite{newick}
(read and write Newick data formatted files),
\texttt{setuptools}~\cite{setuptools} (a packaging library for Python
projects). It is interesting that the prominent R ``Nebraska''
packages belong to the \texttt{tidyverse} system~\cite{tidyverse}, and
thus are co-authored by Hadley Wickham.

To check the results, we included in the tables three popular Python
packages: \texttt{scipy}~\cite{2020SciPy-NMeth}, \texttt{numpy}~\cite{trujillo2022penumbra}, and \texttt{napari}~\cite{napari_contributors_2019}.  We see
that \texttt{scipy} and \texttt{numpy} are in high percentiles of
package by centrality and mentions for all three versions of network
studied. The newer and more specialized package \texttt{napari} is
lower in the ratings, but still is above 50\% percentile.  All this
seems reasonable.  

Similarly we were pleased to note that the analysis identified the \texttt{tifffile}~\cite{christoph_gohlke_2026_19522931} package as part of the Nebraska quadrant and anecdotes and quick inspection of the online materials confirm this as a high quality, widely used package with one person development team, albeit in California and not physically in Nebraska.

The difference between the unweighted and weighted versions of our
analysis, in~\autoref{fig:results} can help us parse different
subtypes of these outliers.  For example, the
\texttt{velvet}~\cite{ZerbinoEtAl2014} package is central solely because it is
mentioned by one highly cited paper (and dozens of less highly cited ones). 
Such characteristics do not make a package central in the unweighted version of the
analysis, since citation counts are ignored there. Central packages under
this representation, therefore, also need to have other packages depend
on them and cannot solely rely on mentions from influential papers.

It is interesting that \texttt{cran} and \texttt{bioconductor}
dominate the ``Nebraska'' quadrant for the unweighted graph, while for
the weighted graph, \texttt{pypi} is overrepresented.  The reason for
this difference may warrant additional research.

Altogether, our paper tries to quantify impact as importance and to
provide a set of metrics behind the intuition.  Of course, this is by itself a
perilous process: there could be various flavors of both impact and
importance, and people may reasonably disagree about the details.  Our
contribution is to propose a certain quantification, which agrees with
our intuitive understanding.

In addition to possible problems with the exact definition of importance,
our work has a number of limitations.
\begin{enumerate}
\item The network analysis is limited by the quality of disambiguation
  and linking in the CZ Software Mentions dataset (see the discussion
  in~\cite{SoftwareMentionsArXiv2022}).  
  Homonyms (different software
  packages with the same or close names) may change the network
  statistics, and preliminary research points to considerable potential for incorrect linking in the CZ Software Mentions dataset due to
  homonyms~\cite{DruskatEtAl2024}.  
  Notably, this affects the
  packages \texttt{velvet} and \texttt{tophat} (see~\autoref{fig:network-viz}a), which 
  have been falsely linked to Python
  packages.
  \texttt{velvet} was linked to a package that provides signal processing and 
  communications algorithms in Python~\cite{velvet}.
  Meanwhile, \textit{all} of the mentions of ``velvet'' point to 
  different versions of algorithms for de novo short-read assembly using de Bruijn graphs
  in genomics, as described in~\cite{ZerbinoBirney2008} and implemented mostly in C~\cite{ZerbinoEtAl2014}.
  \texttt{tophat} was linked to a package that provides a framework for collaborative and multiplayer mobile applications~\cite{tophat}.
  However, over 99~\% of mentions point to different software called \texttt{TopHat}, a splice junction mapper for RNA-Seq reads, implemented mostly in \CPP\cite{trapnell_etal_2009_tophat_discovering_splice,kim_etal_2013_tophat2_accurate_alignment}, and the remainder to the Top-Hat transform algorithm used for baseline subtraction in image processing.
  While the linking errors do not change the centrality values and our interpretation of them -- both rely on disambiguation, not linking --
  they led to mislabeling of different, mostly non-PyPI packages as PyPI packages for the visualization in~\autoref{fig:network-viz}.

  Sometimes an open source package also shares a name with a
  proprietary software (e.\,g.~PRISM, PACE), which leads to confusion and
  incorrect mention counts.

  Nonetheless, we think that this contribution is useful, both as a demonstration of this approach and because our analysis is reproducible. This means that limitations can be addressed in improvements over time. For example, as datasets emerge with improvements to disambiguation these datasets can be incorporated into the analysis and comparisons or updates produced.

\item Our approach is limited to the open-source packages available in one of the chosen ecosystems.  This excludes proprietary
  software like Excel or GraphPad Prism~\cite{swift1997graphpad}, and open-source
  non-package software like Gephi~\cite{ICWSM09154} (a complied application).  In addition, our work excludes software projects such as BLAS~\cite{blackford2002blas}, LAPACK~\cite{anderson1999lapack} or the Linux Kernel itself. These packages are critically important but are often lower-level system dependencies that may not be clearly cited or referenced in an article's text or provided in a software library's dependency list. Without such references or dependencies, it is hard to detect their specific utilization. Without authors including references to lower-level software libraries in their articles, and without explicit dependency on lower-level libraries from mentioned software, we simply cannot include such software in our networks. The number of missing packages from the ecosystems investigated may be estimated from Table~14 of~\cite{SoftwareMentionsArXiv2022}, which estimates the 
  coverage for GitHub at 64.39\%, the coverage for CRAN at 8.36\%, the
  coverage for PyPI at 5.86\%, and the coverage for Bioconductor at
  3.23\%.  GitHub coverage can be used as a proxy for the share of the
  open source software: while some packages---perhaps growing in number---are not hosted by GitHub,
  their fraction remains low~\cite{trujillo2022penumbra}.  If we assume that each package
  exists only in one ecosystem (there are packages present in several,
  but we neglect this effect), the combined coverage by CRAN, PyPI,
  and Bioconductor is 17.45\%, i.e., between a quarter and one third of
  the open source packages.  It is very likely that each of these
  non-package pieces of software do depend on packages in our focus
  ecosystems. For open-source non-package software, it may be feasible
  to identify source repository URLs and then identify packages
  depended on by the software. That would need to be done by directly
  inspecting the source code, possibly using Software Bill of Materials (SBOM) tools~\cite{Hatta2022Nebraska}. On the other hand, some
  non-package software does not have source code available (including
  proprietary GUI software such as SPSS, and cloud-based services).
  Nonetheless, these pieces of code may very well be based on packages
  in our focal ecosystem (especially if the latter use non-restrictive
  licenses such as BSD, MIT).  In the longer term, there is a possibility
  that the requirement to provide Software Bills of Materials (SBOMs) for U.S. federal government
  purchasers will give insight into the packages on which proprietary
  software relies~\cite{howison_support_2024}.
\item In our dependency network, we always used the latest
  version of any package.  In reality, dependencies change between the
  versions.  A study of development logs often reveals messages like
  ``deleted the dependence on XYZ'' or ``added the dependence on
  ABC''.  These changes often make the work of package management
  software very difficult.  A software dependency graph is a living
  network, constantly evolving, with links added and deleted.  In this
  study, we effectively collapse the time, making an (imperfect) snapshot of the
  long movie.  

  A future improvement would be to use each package ecosystem's
  specific dependency resolution algorithm to compute the full
  transitive dependency tree for each mentioned software package.
  An
  even further extension would be to attempt to do this for the
  version of the package most likely used by a specific publication,
  based on aligning publication date with the current version used at
  that time.
\item Our method captures the off-the-shelf software packages used for
  each article.  It does not capture \emph{ad hoc} software written
  specifically for the given paper, which may be contained in the code
  accompanying the papers.  This code may load software libraries and
  thus add software dependencies that are absent from our graph.
\item The dependencies reflected in the package metadata do not
  necessarily reflect the actual usage of the software.  A library
  may be loaded, but not used for the actual execution by
  a given paper.  A more reliable method would be to reproduce the
  analysis done in each paper and capture the actual library
  calls---which is probably prohibitively difficult for the number of
  papers covered.
\item Many Python and R packages ultimately depend on libraries
  outside their ecosystems (most often C and Fortran libraries).
  These dependencies are not easily captured by the package management
  infrastructure and thus are outside of our analysis.
\item Our dependency network does not include the concept of
  alternatives, when package~$A$ depends on the functionality that can
  be provided either by package $B_1$, or package $B_2$.  Some package
  managers like Debian's \textsl{apt} have the possibility of specifying
  alternatives, but this is not a common feature of package managers.  
\item Different communities place different emphasis on infrastructure
  elements such as tests.  This may lead to the
  over-representation of testing infrastructure packages in some cases,
  for example, in the Python ecosystem.  We do not try to attempt to decide
  which elements of the infrastructure are ``actually important'';
  instead, we rely on the judgment of the community of users and
  developers while recognizing that such judgment may be different
  across the landscape.
\item Our current approach does not filter or subset the network into different time periods (i.e. by year) and as such, some of the software identified as central to our network such as \texttt{velvet} and \texttt{tophat} are perhaps historical artifacts that have not been updated in many years. These examples demonstrate that future analysis of software dependency networks should consider temporal data to understand how particular software projects (and more broadly, computational methods) rise in adoption and then decline as new methods become available.
\item There are different software package repositories and ecosystems 
  for Python and R with greatly overlapping dependency graphs.  We used
  PyPI, CRAN and Bioconductor for this study.  Alternative repositories 
  that provide packages for Python and R, such as Conda channels 
  (conda-forge, bioconda, etc.), are also worth investigating as there may be some software projects distributed solely through Conda channels. Conda provides alternate packaging standards from PyPI and CRAN and as such, over a large distribution, the same set of packages sourced from PyPI may have a slightly different dependency graph than one sourced from Conda. Conda also has more cross-language dependencies listed and may allow deeper traversal to find infrastructural libraries like BLAS; other improvements could include using PyPI Wheels or HomeBrew dependency systems. 
  
\end{enumerate}

\section{Conclusions and future work}
\label{sec:concl}

In this work, we investigated the network of software packages used in
the biomedical literature available through PubMed Central.  We
demonstrated that these packages follow the structure of ``Stokes'
diagram'' (famous for "Pasteur's quadrant"), with some packages highly
visible to the end users, and some packages less visible, but
important in the network due to their dependencies.  We discussed the
use of Katz centrality in discovering the important packages and found
examples of such packages in the biomedical field. 

These findings and insights, as well as the underlying methodology, can be used to identify 
critical, but low-visibility, open-source scientific software in need of targeted funding and support.

Of course, there are many ways to extend this work.  It would be
interesting, for instance, to analyze common workflows for different disciplines,
perhaps using co-occurrences of mentions, and map them into the
dependency network.  This might help to discover packages important for
specific sub-fields of biomedical sciences.  Similarly, adding temporal
dependencies to our graph may help to discover and predict 
development trends.
Additionally, individual dependencies may differ in the actual impact that they have on the dependent software (i.e., a dependency that is only minimally utilized by the downstream software itself may be given less weight).
Developing and applying a metric that describes these differences in impact would make it possible to refine our approach to quantifying the importance of research software dependencies.

\bibliography{dependencies}

\end{document}